\documentclass[reprint,amssymb,amsmath, aps, prl, cha, twocolumn]{revtex4-1}
\usepackage{graphicx}
\usepackage[caption = false]{subfig}
\usepackage{color}
\newcommand{\comment}[1]{}
\usepackage{epsfig}
\usepackage{ifpdf}
\usepackage{url}%
\usepackage{multirow}
\usepackage{float}
\usepackage{bm}
\usepackage[colorlinks=true,linkcolor=blue]{hyperref}%
\usepackage{cleveref}
\expandafter\ifx\csname package@font\endcsname\relax\else
\expandafter\expandafter
\expandafter\usepackage
\expandafter\expandafter
\expandafter{\csname package@font\endcsname}%

\usepackage{subcaption}
\usepackage{ragged2e}
\DeclareCaptionJustification{justified}{\justifying}
\fi
\begin{document}
\title{Square lattice formation in a monodisperse complex plasma  }%
\author{Swarnima Singh}%
\email{sswarnima239@gmail.com}
\author{P. Bandyopadhyay}
\author{Krishan Kumar}
\author{A. Sen}
\affiliation{Institute For Plasma Research, A CI of Homi Bhabha National Institute, Bhat, Gandhinagar-382428, India}
\date{\today}
\begin{abstract} 
  We present the first observations of a square lattice formation  in a monodisperse complex plasma system - a configurational transition phenomenon that has long been an experimental challenge in the field. The experiments are conducted in a tabletop L-shaped Dusty Plasma Experimental (DPEx-II) device in a DC glow discharge Argon plasma environment. By a careful control of the  vertical potential confining the charged particles as well as the strength of the ion wake charge interactions with the dust particles we are able to steer the system towards a crystalline phase that exhibits a square lattice configuration. The transition occurs when the vertical confinement strength is slightly reduced below a critical value leading to a  buckling of the monodisperse hexagonal 2D dust crystal to form a narrowly separated bilayer state (a quasi-2D state).  Some theoretical insights into the transition process are provided through Molecular Dynamics (MD) simulations carried out for the parameters relevant to our experiment. 

\end{abstract}
\maketitle
\section{Introduction}\label{sec:intro}
Two dimensional (2D) materials are of great scientific interest not only because of their important practical applications but also due to the fundamental differences in their intrinsic physical characteristics as compared to three dimensional solids  \cite{Zeng2018,Smith2020,wang2018}. The differences are particularly prominent in the nature of their respective phase transitions (e.g. melting transitions) and also transitions leading to configurational changes in the lattice geometry \cite{duerloo2014}.   The latter have a direct impact on the mechanical, optical, chemical and electrical properties of the material and hence a great deal of past studies have investigated the geometrical nature of crystalline lattices in various 2D systems  \cite{Bernal1964,pan1998, Pieranski1983, raizen1992,Bhushan1995,steve1991}. Many of these studies have been carried out in colloidal suspensions where it is possible to do particle resolved studies on a reasonable time scale and using fairly straightforward diagnostics \cite{O_uz_2009,O_uz_2012,neser1997,pansu1983,schmidt1996, Fontecha2007}. \par
More recently dusty plasma (also known as complex plasma) systems have been widely used for the study of two dimensional charged layers that undergo various phase transitions \cite{Melzer1996,Thomas1996,Bin2003}. A dusty plasma is an ensemble of micron-to sub-micron sized particles that are immersed in a plasma of electrons and ions. The dust particles, which become highly negatively charged due their interaction with the plasma, can then go into various phase states including crystalline states due to the strong mutual Coulomb interactions. Typically in an experimental device, the dust particles are levitated by the electrostatic field of an  electrode and attain an equilibrium height when the forces due to the upward electric field and the downward pull of the gravity balance each other. 
 The electric field of the sheath, besides providing confinement to the dust particle, also influences the cloud of ions streaming around the negatively charged dust particle making the cloud highly asymmetric. These so called `ion wakes'  exert an attractive force on the dust particles making their mutual interactions non-reciprocal. Ion wake forces give dusty plasmas a unique character and  distinguish them from other systems like colloidal or conventional plasmas. They play an important role in the phenomena of crystal melting \cite{Couedel2010} as well as in structural transformations \cite{Zampetaki2020}.  Experiments in 3-D dusty plasmas have revealed different structural phases like fcc and bcc \cite{zuric2000, pieper1996} while in a monolayer dust crystal formed from monodisperse particles the predominant observed phase is that of hexagonal structures \cite{Thomas1994}. Square lattice formations have been experimentally observed in binary complex plasmas where two dust species with nearly identical charge to mass ratios are seen, under certain conditions,  to form a narrowly spaced bilayer which exhibits square lattice patterns \cite{huang2019}. For a monodisperse dust layer square lattices have been observed in MD simulations by inducing a buckling in the crystal layer through a manipulation of the confinement potential and wake-field interactions \cite{Zampetaki2020}. However, to date, there has been no experimental demonstration of a square lattice formation in a monodisperse complex plasma.\par
 
A major requirement for transitioning to such a state is the creation of a stable bilayer (a quasi-2D state) from the monolayer hexagonal crystal structure. This can be done through buckling the monolayer by altering the harmonic component of the confining potential as was demonstrated in the MD simulations of Zampetaki {\it et al.} \cite{Zampetaki2020}. Such a technique has been successfully tried out experimentally in different systems like laser cooled trapped ions and colloids. In a complex plasma the presence of wake field interactions promotes vertical pairing of particles between the two layers which become unstable for strong values of the wake field and leads to melting and destruction of the crystal layer. This instability, known as the mode coupling instability (MCI), is a major deterrent for the formation of a square lattice structure in a monodisperse complex plasma. As discussed in \cite{Zampetaki2020}, one needs to suppress the MCI through proper control of the wake field interactions as well as neutral damping to successfully attain the desired structural phase transition. In our experimental device the asymetric electrode configuration has a direct influence on the course of the streaming ions that helps in the stability and ease of formation of the crystalline state \cite{Arumugam_2021}. By preventing the ion wake interactions to achieve high values and through a judicious control of the background pressure as well as the confining potential we are able to transition from an extended hexagonal monolayer crystal to a state with a number of small domains of square lattices. Local bond order parameter and structure static factor are used to distinguish between the different structural phases of the system. The experimental observations are also validated with MD simulation results relevant to our experimental conditions. \par
\section{Experimental Set up and procedure}
\begin{figure}[t!]
\begin{center}
\includegraphics[scale=0.6]{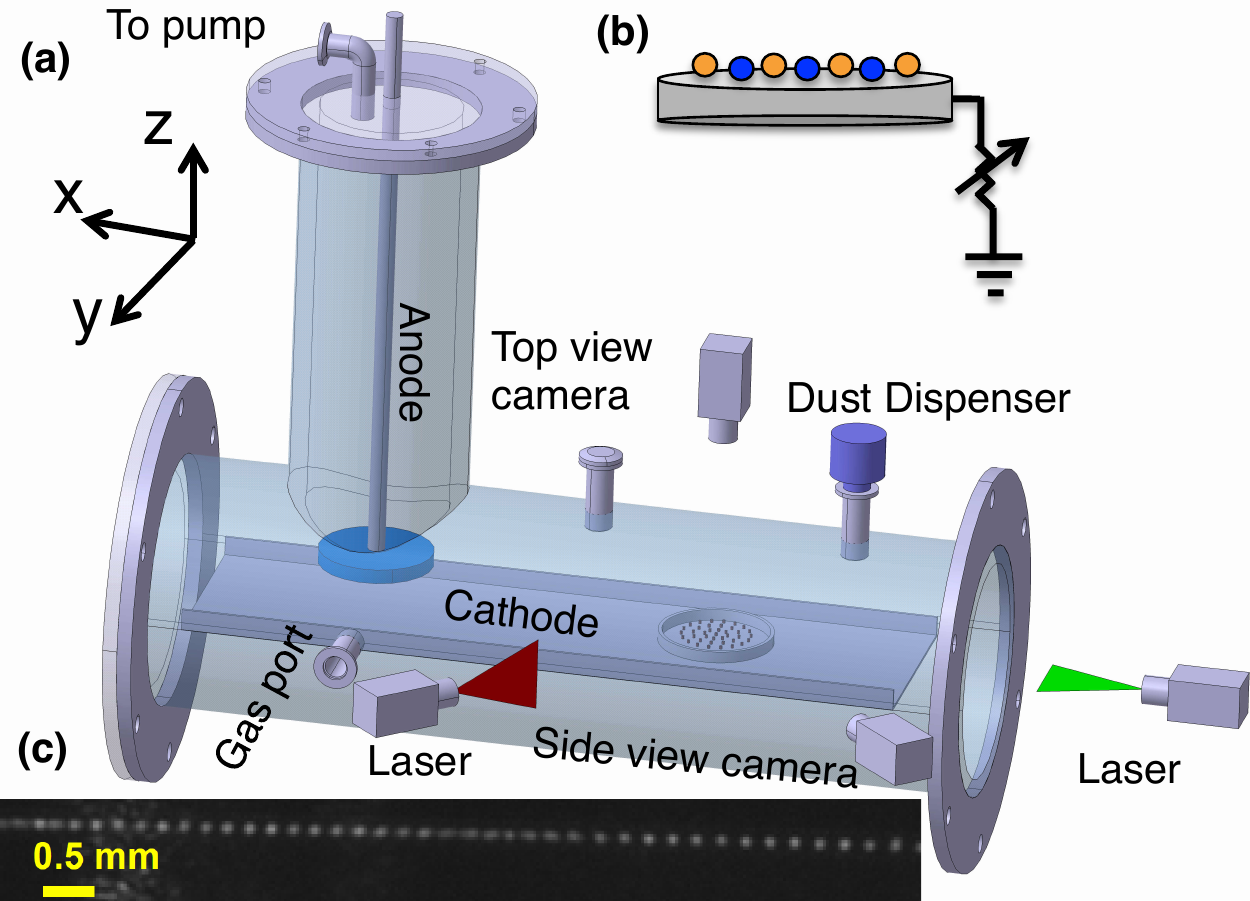}
\caption{(a) The schematic of the experimental set-up, (b) A schematic vertical view of confinement ring with dust particles (c) Side view of dust crystal at P = 4.8~Pa and V = 406~V.}
\label{fig:setup}
\end{center}
\end{figure}

The experiments are conducted in a newly built L-shaped DC dusty plasma experimental (DPEx-II) device as shown in Fig.~\ref{fig:setup}(a) \cite{Arumugam_2021}. A SS ring of diameter 6~cm is placed on the cathode to trap the dust particles horizontally. This ring is kept electrically isolated from the cathode and connected through a potentiometer, as shown in Fig.~\ref{fig:setup}(b). This arrangement allows us to change the confinement strength of the ring by maintaining the same discharge condition.  With the increase in the resistance attached with the ring, the sheath gets squeezed resulting in a change in the confinement of the particles \cite{krishan_2021}.\par

In the present set of experiments, a weakly ionized Argon plasma is initially created between the electrodes using a DC voltage of 435~V at a gas pressure of 4.8~Pa. Monodisperse melamine-formaldehyde (MF) particles of diameter 4.33~$\mu m$ are injected into the plasma to form a dust crystal. Prior to injecting these MF particles into the plasma, their size distribution was checked by examining them under a scanning electron microscope (SEM). The particles were indeed found to be monodisperse with a size distribution spread 
$\sim 1.5\%$ (see Sec.~I of supplemental material \cite{supplement} for further details). Two laser sheet beams of different wavelengths (green and red) are used to illuminate the particles in the horizontal and vertical directions, and the dynamics of the particles are captured with the help of top- and side-view cameras. \par
\begin{figure}[t!]
\includegraphics[scale=0.45]{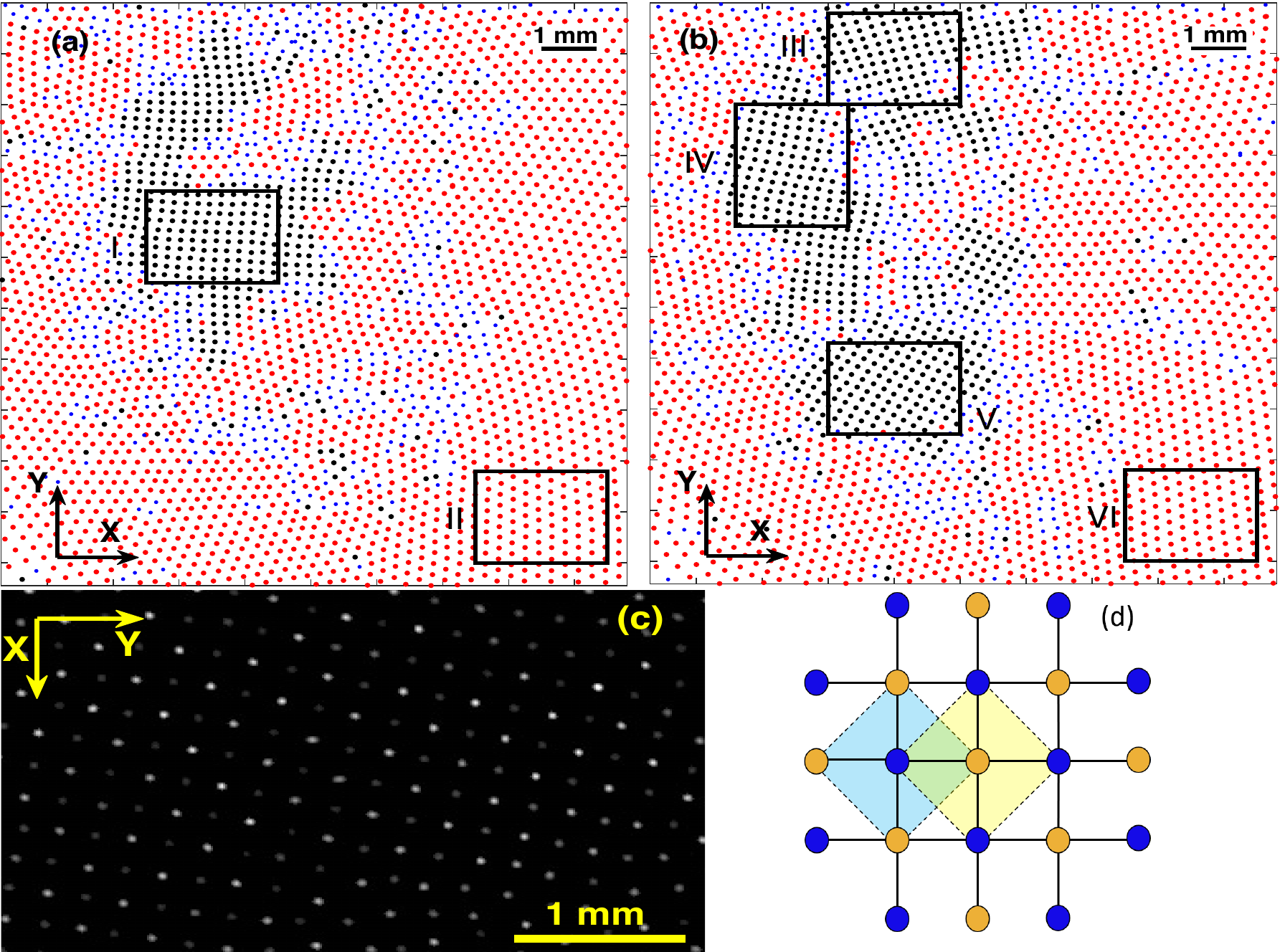}
\caption{Local structural arrangement displayed by the order parameter ($\Psi_n(j)$) of the particles in the monodisperse complex plasma experiment at discharge voltages (a) V = 400~V, and (b) V = 396~V at a constant pressure of P = 4.8~Pa. The particles with red color have $\Psi_6>0.6$, black has $\Psi_8>0.6$, and the blue color corresponds to defects. The rectangles of Fig.~\ref{fig:snapshot}(a) and (b) mark the area having different structural arrangements. (c) Top zoomed-view of region IV of Fig.~\ref{fig:snapshot}(b) shows the  $2\square$ structure in complex plasma observed in our experiments. (d) A typical sketch of unit cells of a bilayer square structure ($2\square$). All the particles in Fig.~\ref{fig:snapshot}(d) are identical and their colors distinguish particles of two different layers. The shaded regions mark the unit cells of the corresponding lattices.}
\label{fig:snapshot}
\end{figure}
\section{Results and discussion}
 A circular monolayer of dust particles  is formed at the discharge voltage of V $\sim$ 435~V and pressure of P $\sim$ 4.8~Pa. In this condition, the resistance of the potentiometer is set at  3~$k\Omega$. The plasma parameters such as plasma density and electron temperature, obtained from  single and a double Langmuir probe measurements, are in the range of $1-2$~x~$10^{15}/m^3$ and $2-3$ eV, respectively, for our discharge conditions \cite{Arumugam_2021}. The confinement strength is altered by adjusting the discharge voltage (V) and keeping the gas pressure (P) constant.  The system continues to display a hexagonal monolayer configuration till  the discharge voltage reaches a value of 406~V, The images taken by the side-view camera (see Fig.~\ref{fig:setup}(c)) essentially show that the dust particles indeed form a monolayer. Fig.~\ref{fig:setup}(c) also clearly indicates that there is no evidence of the formation of dimer or agglomerated particles as these would inevitably levitate at the bottom of the monolayer due to a lower charge to mass ratio.\par
In order to measure the degree of square and hexagonal order in the system, we use the quantities $\Psi_8$ and $\Psi_6$, respectively, where the local bond order parameter $\Psi_n$ with n = 1, 2, 3,.. for each particle is defined as \cite{Bonitz}:
\begin{equation}
\Psi_n(j)=\frac{1}{N_j}\sum^{N_j}_{l=1}\exp[in\Theta_{j,l}]
\end{equation}
where $N_j$ is the number of nearest neighbour of each particle j, calculated form Delaunay triangulation. $\Theta_{j,l}$ is the angle between the nearest-neighbor bond from particle $j
$ to $l$ with respect to x direction.  $\Psi_6$ ($\Psi_8$) is expected to be one for a perfect hexagonal (square) lattice and zero if no hexagonal (square) order is present in the lattice. It is observed that when the confinement strength is decreased further by decreasing the discharge voltage (V = 400~V), a domain with square lattice structures forms at a location (region-I) of the extended dust crystal as shown in Fig.~\ref{fig:snapshot}(a). 
Apart from the visual evidence, the particles in the region-I of Fig.~\ref{fig:snapshot}(a) also show a value of $\Psi_8$ that is  different from the typical hexagonal lattice as seen in the rectangular region-II of Fig.~\ref{fig:snapshot}(a). The static structure factor \cite{Vasilieva2021} also provides a useful signature to distinguish between different lattice arrangements as is discussed in detail in Sec.~II of the supplemental material \cite{supplement}.\par
\begin{figure}[h!]
\begin{center}
\includegraphics[scale=0.53]{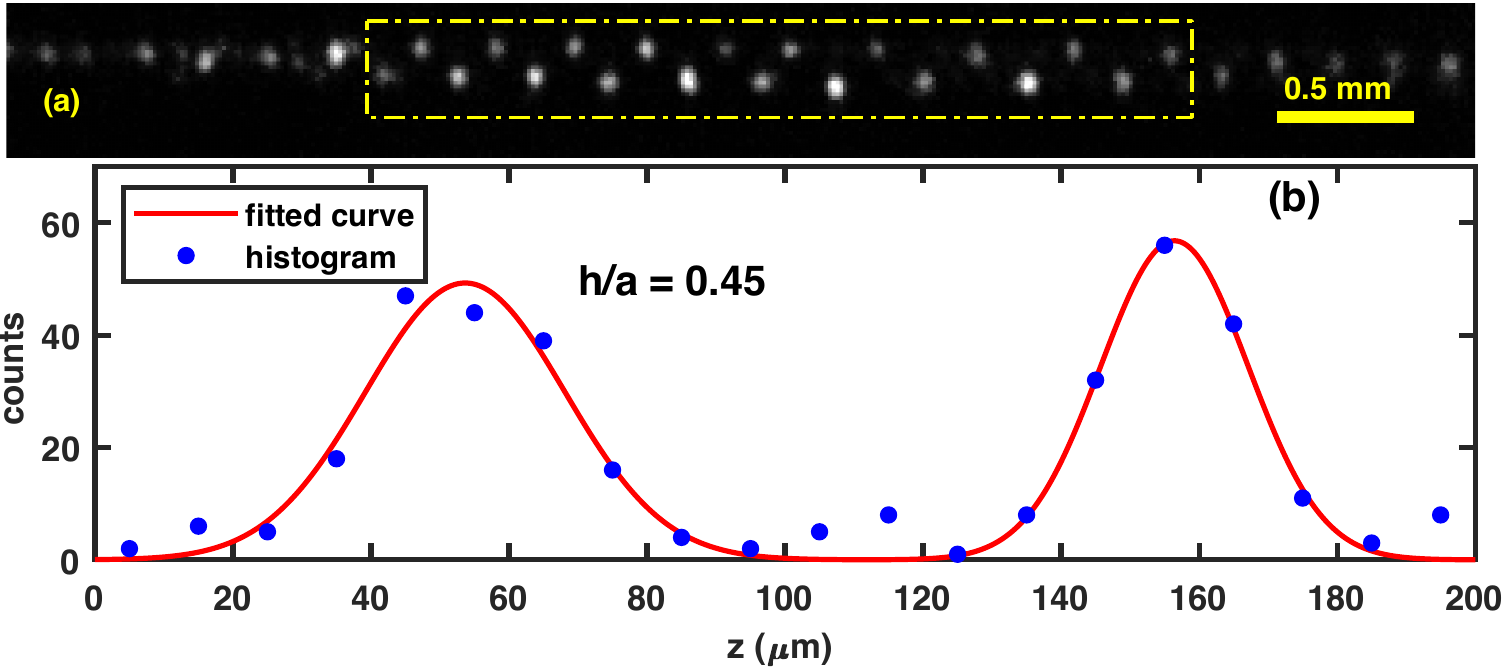}
\caption{(a) Side view of the particles at P = 4.8 Pa and V = 400 V. The particles in the dotted box represents the quasi 2D layer, (b) histogram of particle's position in z-direction corresponding to experimental condition mentioned in (a).}
\label{fig:side}
\end{center}
\end{figure}

 
With a further reduction in the confinement strength (V = 396~V), many square domains with different orientations form (see regions III, IV and V of  Fig.~\ref{fig:snapshot}(b)). These square lattice structures are found to be very stable in time as long as the discharge condition is not altered. Fig.~\ref{fig:snapshot}(c) shows a top zoomed-view of the bilayer with a square structure (represented by the notation  $2\square$) formed in region-IV of Fig.~\ref{fig:snapshot}(b).  The position of the maximum intensity of the laser is set to the topmost layer of dust crystal. The difference in the intensity of the particles clearly shows that the particles do not reside in the same layer. It is to be noted that the interlayer width is much smaller than the interparticle distance ($a$) in the horizontal layer. This separation is well resolved since the width of the laser thickness is also smaller than the $a$. The arrangement of the particles is the same as shown in the sketch of a regular bilayer square structure as shown in Fig.~\ref{fig:snapshot}(d). The quasi 2D structure of the square lattice can also be ascertained from the side view of the dust system as shown in Fig.~\ref{fig:side}(a). It can be seen that the positions of the particles in the bottom layer of the quasi-system are laterally shifted with respect to those of the top layer. In other words, they do not form  vertical pairs with particles in the top layer \cite{Nosenko2014}. To examine the $\Delta z$ distribution of this quasi system, the particles are tracked along the vertical direction (z) for several frames and used to construct a distribution function that is shown in Fig.~\ref{fig:side}(b). The $\Delta z$ distribution essentially implies that the particles indeed form a bilayer system at $V$ = 400~V with the two distinct peaks in Fig.~\ref{fig:side}(b)  indicating that the two layers are separated by a distance of $\tilde{h}$ = $h/a$ = 0.45 i.e., 102 $\mu$m in the z-direction. The variation of $\tilde{h}$ with the discharge voltage is discussed later in this paper.\par
The unique electrode configuration \cite{Arumugam_2021} used in our DC discharge has the effect of moderating the dust-wake interaction. This configuration allows the ions to move towards the cathode with a higher Mach number (M) and leads to a weaker wake potential as compared to those commonly found in RF discharges \cite{Ludwig_2012, Kompaneets2016,hou2001}. Typically in RF discharges, the ion-wake potential is of the order of $\sim 20$ mv \cite{Ludwig_2012}. By contrast, the estimated wake potential for our discharge condition (P=4.8~Pa and $V=400$~V) is about  $\sim 2.5$~mV for M=4.5.  A detailed calculation of the wake potential has been provided in Sec.~III of the supplemental material \cite{supplement}. For a higher Mach number, the wake gets stretched in the streaming direction while the potential\rq{s} peak height goes down and thus ultimately decreases the extent of the area of the wake interaction \cite{Ludwig_2012,ion_wake}. Due to this wake suppression, the vertical pairing in our experiments become insignificant, which ultimately paves the way for the formation of the square lattice in the complex plasma. In this situation, when the external confinement is decreased below a critical value via decreasing the discharge voltage, the out-of-plane mode of dust lattice wave becomes purely imaginary \cite{Zampetaki2020}. It induces a transverse instability in the system, making the structure unstable. The system tries to avoid this situation by transitioning to a different lattice structure that has  better  stability. This results in the structural phase transition in the dust system, where a monolayer hexagonal structure buckles to a bilayer square structure \cite{Zampetaki2020}. \par
\begin{figure}[t!]
\begin{center}
\includegraphics[scale=0.7]{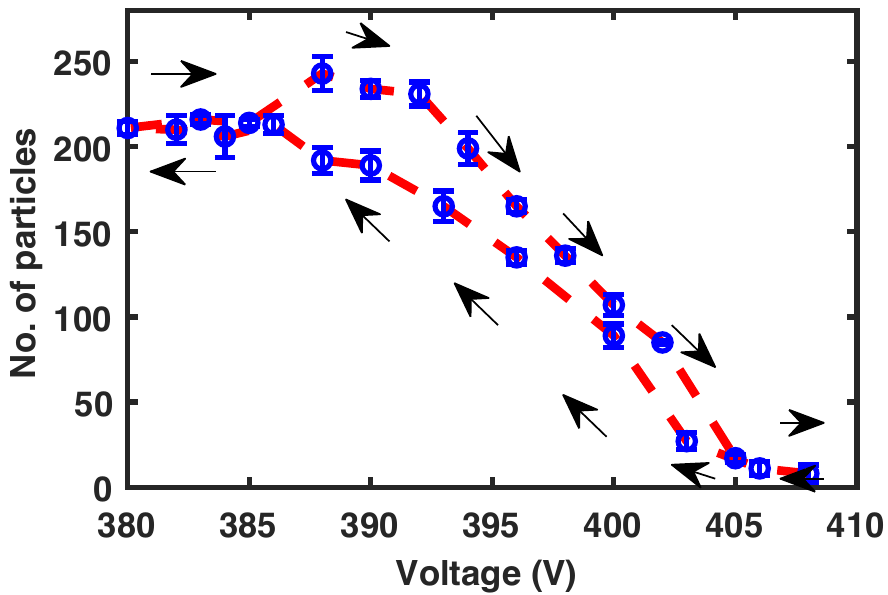}
\caption{A hysteresis behaviour in the system with discharge voltage as a control parameter. The Y-axis represents the number of particles in the region of interest having $\Psi_8 >$ 0.6.}
\label{fig:hyst}
\end{center}
\end{figure}
To examine the presence of any hysteresis phenomenon in the system, the experiments are repeated by varying $V$ in small steps and keeping the other parameters constant. For a given value of $V$, we have estimated $\Psi_8$ for all the particles and considered those particles whose $\Psi_8$ value is more than 0.6. At a higher discharge voltage of $V=408$~V, a negligible number of particles show the square symmetry in the region of interest of the crystal. The number of particles, which possess $\Psi_8 > 0.6$ increases with the decrease of discharge voltages and approximately 210 particles possess $\Psi_8 > 0.6$ at $V\sim385$~V, which remains almost constant even if the discharge voltage is decreased further. After attaining  $V=380$~V, the discharge voltage is reversed and it is found that more number of particles now possess $\Psi_8 > 0.6$ at all the discharge voltages compared to the case when the discharge voltage is decreased, which reveals a clear hysteresis behaviour in our system. The hysteresis curve is shown in Fig~\ref{fig:hyst}. The hysteresis behaviour in the present case is associated with the  structural transition from a hexagonal lattice to a square lattice \cite{Zampetaki2020}.\par
\begin{figure}[t!]
\begin{center}
\includegraphics[scale=0.80]{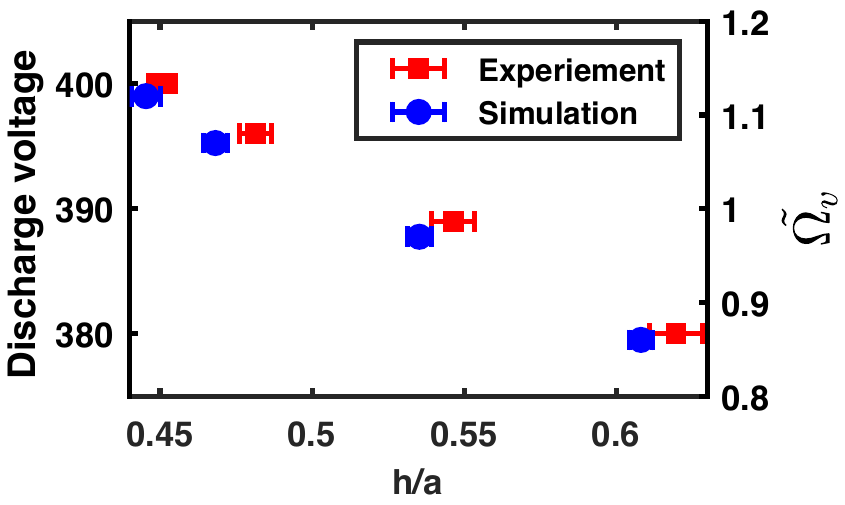}
\caption{ The evolution of the average interlayer separation (normalised with  $a$)  with discharge voltage. For comparison, we also show the normalised average interlayer separation at different confinement frequency ($\tilde{\Omega}_v$) found in the MD simulation. }
\label{fig:z}
\end{center}
\end{figure}

As theoretical support for our experimental results, we have carried out Molecular Dynamic (MD)) simulations using Langevin dynamics  and in which we have included a point-wake model \cite{rocker2014} to represent the experimental conditions. The MD simulations are performed using an open source code, LAMMPS \cite{PLIMPTON19951} by incorporating the necessary forces \cite{Zhdanov2009, Tomme2000} . The parameters used in the simulation are chosen to be close to the experimental values. The details are given in the Sec.~IV of the supplemental material \cite{supplement}.
\begin{figure}[h!]
\begin{center}
\includegraphics[scale=0.38]{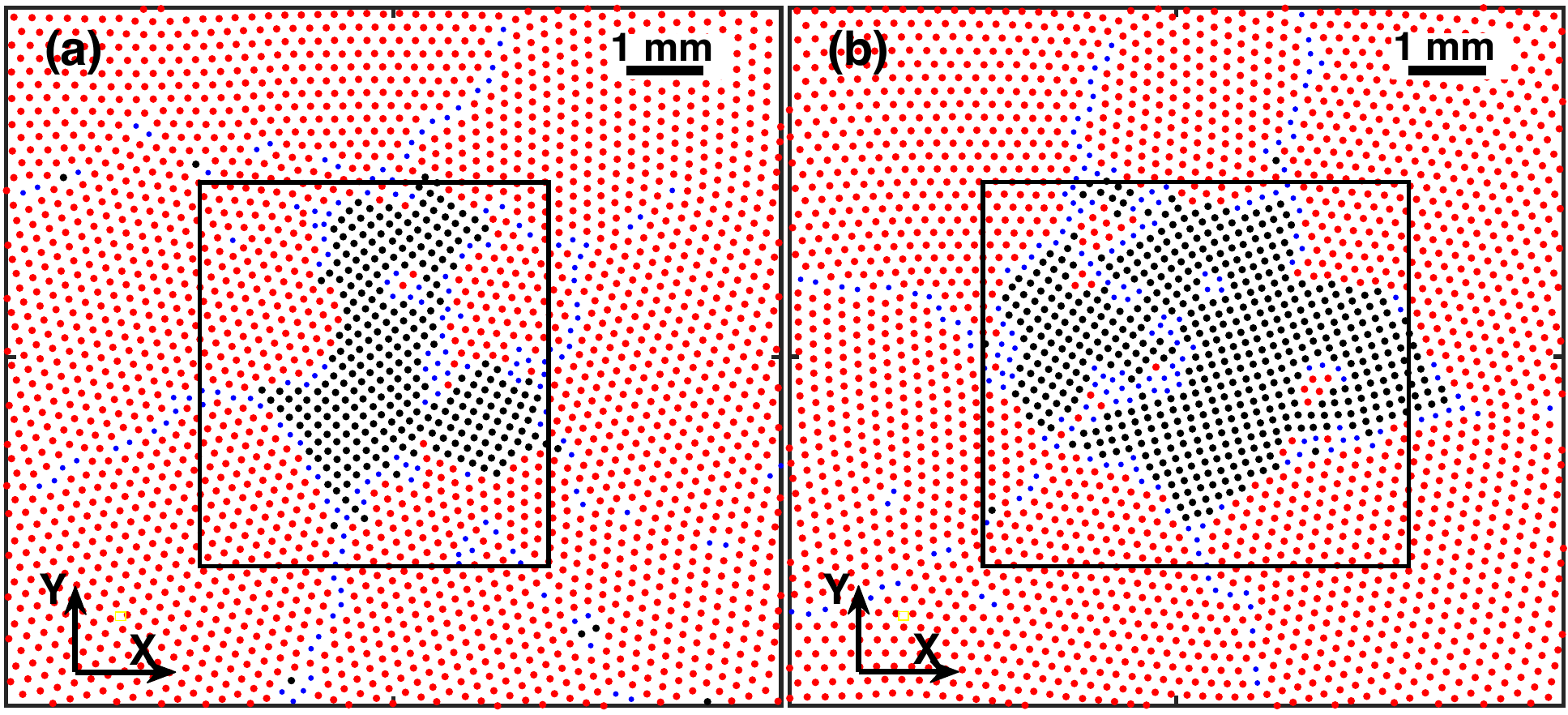}
\caption{Snapshot of particle positions resulting from MD simulations for experimentally relevant parameters at $\tilde{\Omega}_h$ = $0.026$ (a)  $\tilde{\Omega}_v$ = $1.07$ and (b)  $\tilde{\Omega}_v$ = $0.97$. Black color represents particle having $\Psi_8>0.6$, red particles have $\Psi_6>0.6$ and blue particles are defects. The parameters are $\kappa=5.7$ and $\Gamma = 3563$. }
\label{fig:con_simu}
\end{center}
\end{figure}
As shown in Fig.~\ref{fig:con_simu}(a), if one decreases the vertical confinement frequency $\tilde{\Omega}_v$, some regions in the hexagonal monolayer system start showing  square structures at a value of $\tilde{\Omega}_v=1.07$. Several new domains of square lattice are formed with a further reduction of $\tilde{\Omega}_v$, thus increasing the extent of square lattice in the system (see Fig.~\ref{fig:con_simu}(b)). In the simulations, the square configuration is accompanied by an increase of the interlayer separation $\tilde{h}$. The evolution of $\tilde{h}$ with the  confinement frequency ($\tilde{\Omega}_{v}$)  is shown in Fig~\ref{fig:z}.  The figure clearly shows that the separation between the two layers increases with a decrease in the $\tilde{\Omega}_{v}$. Likewise, we have found a similar trend in the separation of the two layers when the discharge voltage is decreased in the experiments. Furthermore, the experimental values for the $\tilde{h}$ are in excellent agreement with those obtained  by the simulation. These measurements are also consistent with the simulation results given by Zampetaki \textit{et al}. \cite{Zampetaki2020}.
\par
 \section{Conclusion}
 In this letter, we have demonstrated the first-ever experimental observation of square lattice formation in a monodisperse complex plasma crystal. Bilayer domains containing square lattice structures coexist  with the initial larger monolayer with hexagonal structures  for a given confinement potential below a critical value.  MD simulations carried out for similar physical parameter values reproduce the dominant features of the experimental observations. For a particular value of wake charge, we obtain a threshold vertical confinement frequency below which the hexagonal monolayer bifurcates into several square lattice domains. In a recent study, Zampetaki \textit{et al.} have theoretically modelled the  transition in a non-reciprocal plasma crystal as a multi-step process where one passes sequentially from a single hexagonal layer ($1\Delta$) to a bilayer square ($2\square$) via an intermediate state consisting of a hexagonal bilayer structure with a doubly occupied bottom layer (21) \citep{Zampetaki2020}. However, we do not observe such an  intermediate state either in our experiment or in our simulations. Thus there still remain open issues in this interesting topic.  We believe our findings can be the basis for  future experimental, theoretical, and simulation studies on buckling induced phase transitions in 1D and 2D non-reciprocal systems.  This can not only provide rich insights into the behaviour of strongly coupled two-dimensional complex plasma systems but also help to develop useful linkages with the dynamics of a wider set of soft condensed matter systems.
\begin{acknowledgments}
A.S. is thankful to the Indian National Science Academy (INSA) for the INSA Honorary Scientist position. All the simulation work reported here has been performed on Antya cluster at the  Institute for Plasma Research, Gandhinagar,
India.
\end{acknowledgments}
%
\end{document}